\documentclass{article}
\usepackage{arxiv}
\usepackage[utf8]{inputenc} 
\usepackage[T1]{fontenc}    
\usepackage{hyperref}       
\usepackage{url}            
\usepackage{booktabs,dcolumn,multirow,tabularx}       
\usepackage{amsfonts}       
\usepackage{nicefrac}       
\usepackage{microtype}      
\usepackage[numbers]{natbib}         
\usepackage{lipsum}
\usepackage{graphicx}
\usepackage{siunitx}
\usepackage{wrapfig}
\usepackage{subcaption}
\usepackage{siunitx}

\title{A general machine learning model of aluminosilicate melt viscosity and its application to the surface properties of dry lava planets}

\author{
 Charles Le Losq \\
  Université Paris Cité\\
  Institut de physique du globe de Paris\\
  CNRS-UMR 7154\\
  Institut Universitaire de France\\
  Paris, France \\
  \texttt{Corresponding author: lelosq@ipgp.fr} \\
   \And 
 Clément Ferraina \\
  Université Paris Cité\\
  Institut de physique du globe de Paris\\
  CNRS-UMR 7154\\
  Paris, France \\
  \texttt{ferraina@ipgp.fr} \\
  \And 
 Paolo A. Sossi \\
  Institute of Geochemistry and Petrology \\
  ETH Zürich\\
  Zürich, Switzerland \\
  \texttt{paolo.sossi@eaps.ethz.ch} \\
  \And 
 Charles-Édouard Boukaré \\
  York University\\
  York, Canada \\
  \texttt{boukare@yorku.ca} \\
}

\begin{document}
\maketitle
\begin{abstract}
Ultra-short-period exoplanets like K2-141 b likely have magma oceans on their dayside, which play a critical role in redistributing heat within the planet. This could lead to a warm nightside surface, measurable by the James Webb Space Telescope, offering insights into the planet's structure. Accurate models of properties like viscosity, which can vary by orders of magnitude, are essential for such studies.

We present a new model for predicting molten magma viscosity, applicable in diverse scenarios, including magma oceans on lava planets. Using a database of 28,898 viscosity measurements on phospho-alumino-silicate melts, spanning superliquidus to undercooled temperatures and pressures up to 30 GPa, we trained a greybox artificial neural network, refined by a Gaussian process. This model achieves high predictive accuracy (RMSE $\approx 0.4 \log_{10}$ Pa$\cdot$s) and can handle compositions from SiO$_2$ to multicomponent magmatic and industrial glasses, accounting for pressure effects up to 30 GPa for compositions such as peridotite.

Applying this model, we calculated the viscosity of K2-141 b's magma ocean under different compositions. Phase diagram calculations suggest that the dayside is fully molten, with extreme temperatures primarily controlling viscosity. A tenuous atmosphere (0.1 bar) might exist around a 40° radius from the substellar point. At higher longitudes, atmospheric pressure drops, and by 90°, magma viscosity rapidly increases as solidification occurs. The nightside surface is likely solid, but previously estimated surface temperatures above 400 K imply a partly molten mantle, feeding geothermal flux through vertical convection.

\end{abstract}


\section{Introduction}

Magma oceans are recognised as key phenomena shaping the internal structure and secondary atmospheres of rocky planets~\cite{elkins-tanton2012,sossi2020,gaillard2022}. While primarily theoretical, spectroscopic observations of ultra-short period (USP) exoplanets~\cite[e.g. 55 Cnc e and K2-141b; see for a review][]{chao2021} may help shed light on their nature and longevity. Indeed, these "lava planets" are thought to host vast magma oceans on their dayside because of intense stellar irradiation producing dayside temperatures well above 2000 K~\cite{chao2021}. Outgassing from these magma oceans is expected to influence the planet's atmospheric composition, introducing elements such as H, C, N, Si, Fe, Na, and K~\citep[e.g.][]{charnoz2023}. Therefore, spectroscopic observations of the atmosphere of lava planets may provide indirect data documenting the presence and composition of magma oceans. To do so requires a sound understanding of the interactions between the molten mantle and the atmosphere of a lava planet. Combining fluid dynamics simulations~\citep[e.g., ][]{meier2023,boukare2023} with outgassing and atmosphere models~\cite[e.g., ][]{charnoz2023,falco2024} are promising avenues by which this can be achieved. However, despite the arrival of new spectroscopic data ~\cite[e.g., see][]{hu2024} via MIRI and nirSpec on the James Webb Space Telescope, interpreting spectroscopic data in a geodynamic and geochemical context remain challenging.

This is because information as to the key physical parameters that dictate the behaviour of magma oceans is currently missing. Chief among these is magma viscosity ($\eta$, Pa·s). It controls melt mobility and elemental diffusion timescales, and thus the vigor of thermal convection and magma outgassing. Magma viscosity varies non-linearly over more than 15 orders of magnitude with temperature, melt chemical composition (including volatile concentration), crystal and bubble fractions, and pressure~\cite[see for a review][]{russell2022}, posing a significant challenge for accurate calculations.

Models have been developed to calculate the viscosity of crystal- and bubble-free silicate melts~\cite{russell2022}, including empirical models~\cite[e.g.][]{shaw1972a,giordano2008}, high accuracy models for specific compositions~\cite[e.g.][]{romine2015,russell2024}, models based on theoretical frameworks~\cite[e.g.][]{lelosq2017,starodub2019}, and machine learning (ML)~\cite[e.g.][]{lelosq2021,langhammer2022,cassar2023,lelosq2023b}. Regardless of the method, most existing models focus on predicting the viscosity of crystal-free melts at room pressure over a restrained compositional domain (e.g. industrial or geologic compositions).

No existing model can predict magma viscosity across the wide range of melt compositions ($X$), temperatures ($T$), and pressures ($P$) found on USP planets, where compositions could range from Earth-like to refractory or even carbon-rich ~\cite{leger2011,madhusudhan2012,zieba2022}, and where temperatures vary greatly between day and night sides ~\cite[e.g.][]{leger2011,nguyen2020}. Besides, $P$ may exceed those found in the Earth given that the masses of super-Earths are up to 10 times as massive as Earth. Predicting the properties of molten rocks under such conditions require a new generation of models that leverage as much as possible the existing data on alumino-silicate (\textit{sensus latto}) compositions.

In this study, we present a new database of viscosity measurements as a function of $T$ and $P$ for melts with diverse compositions in the system SiO$_2$-FeO-Fe$_2$O$_3$-TiO$_2$-Al$_2$O$_3$-MnO-MgO-CaO-Na$_2$O-K$_2$O-P$_2$O$_5$-H$_2$O. We benchmark ML models to predict magma viscosity and propose a new model combining a Gaussian process with an artificial neural network to calculate melt viscosity and associated uncertainties. We then apply this model for the calculation of the rheology of the magma ocean at the surface of K2-141 b~\cite{malavolta2018,barragan2018}, a dry (i.e. no thick atmosphere) USP planet. We further assess the impact of internal temperature on the nightside surface, potentially allowing inferences about the interior state of dry lava planets from nightside temperature measurements.

\section{Materials and Methods}
\label{sec:headings}

In this study, we aim at performing the following task. Given $T$, $P$ and $X$, a ML model $G$ containing a set of adjustable parameters $\psi$ will calculate $\log_{10} \eta$ as

\begin{equation}
    \log_{10} \eta = G(T,P,X,\psi)
    \label{eq:funcform}
\end{equation}

The adjustable parameters $\psi$ will be learned during a training phase. It consists in solving a least-square regression problem: given $\eta$ measurements at known $T$, $P$ and $X$, parameters $\psi$ are adjusted via gradient descent to minimize the root-mean square error (RMSE) between $\eta$ predictions and measurements. To implement such a ML model, we need a dataset that will be prepared prior to model training. Different ML models can be used to solve the present problem. We will benchmark the performance of a few selected models to determine which one may be most appropriate for the task at hand. Those steps are described in the following sub-sections.

\subsection{Dataset}
As a starting point, we used the database of i-Melt~\cite{lelosq2023b}, which includes $\eta$ versus $T$ measurements for 790 melt compositions in the system Na$_2$O-K$_2$O-CaO-MgO-Al$_2$O$_3$-SiO$_2$. This database was enhanced by a survey of the literature, targeting multi-component compositions including the additional elements FeO, Fe$_2$O$_3$, TiO$_2$, MnO, P$_2$O$_5$ and H$_2$O. As of 25/09/2024, this database comprises 15,440 viscosity measurements in the range 10$^{-3}$ - 10$^{15}$ Pa·s for 2,155 melt compositions different by at least 0.1 mol\% at room pressure, and 1,227 viscosity measurements for 243 melt compositions at high pressures, up to 30 GPa. It represents the compilations of data from 245 publications. When available, the fractions of iron as ferrous and ferric were compiled. When not available, those were calculated using the Borisov model \citep{borisov2018}; in the case no oxygen fugacity details were provided in the publications, we assumed that melt viscosities were measured in air. The following features thus are available in the database: the molar proportions of SiO$_2$, TiO$_2$, Al$_2$O$_3$, FeO, Fe$_2$O$_3$, MnO, MgO, CaO, Na$_2$O, K$_2$O, P$_2$O$_5$, H$_2$O, pressure in GPa, temperature in Kelvin, and viscosity in log$_{10}$ Pa·s. This hand-held database is available at~\cite{ferraina2024}.

\begin{figure}[ht]
    \centering
    \includegraphics[width=\textwidth]{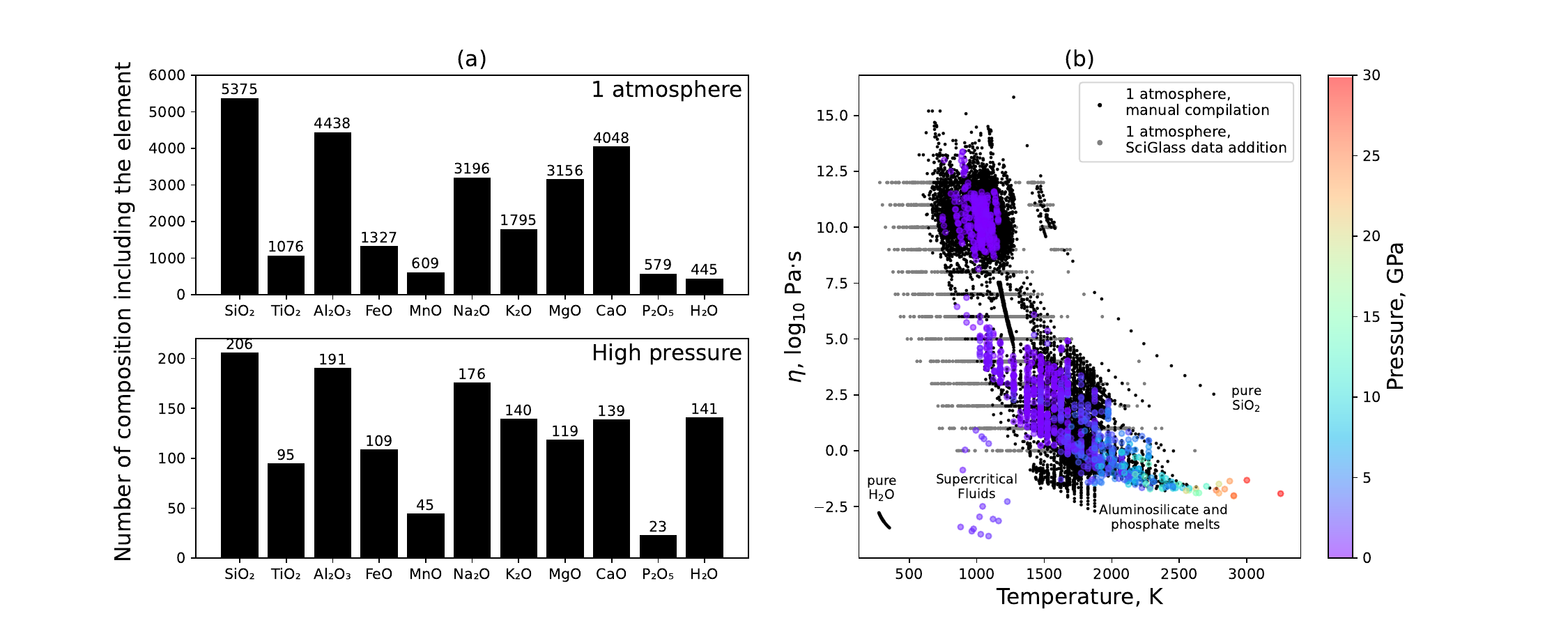}
    \caption{(a) Bar plots showing the number of compositions including a particular oxide component in the database. (b) Viscosity versus temperature diagram showing all data from the full database. Data from the SciGlass database are shown in grey. Colored symbols indicate high pressure data.}
    \label{fig:dataset}
\end{figure}

We enhanced this hand-held database by getting additional data from SciGlass using GlassPy~\cite{cassar2020}, extracting 12,231 data points from 3,591 compositions (different by at least 0.1 mol\%) that do not appear in the hand-held database. Figure~\ref{fig:dataset} provides a visual summary of the entire database, in which the number of compositions including the various elements are visible, together with the viscosity versus temperature and pressure coverage.

\subsection{Data Preparation}

We first separated the dataset in three training (80\%), validation (10\%) and testing (10\%) subsets (see Appendix A for details). The training subset is used for training ML algorithms and the validation subset for tuning their hyperparameters (see below). The testing subset contains data not used during training and hyperparameter tuning, such that it allows evaluating the predictive uncertainties of the models.

$T$, $P$ and the fractions of SiO$_2$, TiO$_2$, Al$_2$O$_3$, FeO, Fe$_2$O$_3$, MnO, MgO, CaO, Na$_2$O, K$_2$O, P$_2$O$_5$, and H$_2$O are features that are provided as inputs to the ML models (eq.~\ref{eq:funcform}). In an attempt to decrease further the predictive errors of the ML models, we tested adding new features (such as the glass optical basicity) but no significant improvements were observed. Therefore, for the sack of simplicity, we only used the features listed above.

A final data preparation step entailed scaling the data. For the four Scikit Learn models (see below), we used a standard scaler, and for other custom models we used a custom scaling where features are 1000./T (K), P/30. (GPa), and oxide fractions.

\subsection{Machine learning models}

Different ML models are appropriate for performing the task described by equation \ref{eq:funcform}. In this work, we benchmark six models. Four are archetypal supervised regression models easy to implement using SciKit Learn~\cite{pedregosa2011}: 
\begin{itemize}
    \item LR is a linear model, used as a baseline;
    \item SVM is a support vector machine model;
    \item RF is a random forest algorithm;
    \item ANN is a feed-forward artificial neural network model.
\end{itemize} 
These models are described in machine learning textbooks~\cite{goodfellow2016,murphy2012}, to which we refer the readers for exhaustive descriptions. Briefly, the LR model establishes a linear relationship between inputs ($T$, $P$ and $X$) and outputs ($\log_{10} \eta$). SVM performs the same task but after projecting the data in a hyperspace using a mathematical trick called the kernel trick. Non-linear relationships become linear on such a hyperplane, such that SVM has the ability to reproduce non-linear relationships between inputs and outputs. RF is an ensemble learning method that operates by constructing a multitude of decision trees at training time and outputting the average prediction of the individual trees at inference time. It is particularly suited to help in addressing the non-linearity and complex interactions between variables. Finally, the ANN model consists of interconnected nodes (a.k.a.~neurons) that `react' differently to their inputs, thanks to non-linear activation functions (e.g., tanh) used to modulate the response of each node to its inputs. This is a foundation structure used in deep learning models, including those for silicate melt viscosity predictions~\cite{cassar2023,langhammer2021,lelosq2021,lelosq2023b}.

Then, we test a greybox artificial neural network (hereafter called Greybox ANN) model that embeds the Vogel-Tamman-Fulcher equation~\cite{russell2022}:
\begin{equation}
    \log \eta = A + \frac{B}{T-C} \,,
    \label{eq:TVF}
\end{equation}
with $T$ the temperature in K, $A$ a common adjustable parameter~\cite{russell2022}, and $B$ and $C$ adjustable parameters that depend on melt composition $X$ and pressure $P$. The artificial neural network predicts $A$, $B$ and $C$ given $P$ and $X$, then equation~\ref{eq:TVF} is used to calculate $\log_{10} \eta$ as a function of $T$. The use of equation~\ref{eq:TVF} constrains the $\log_{10} \eta$ versus $T$ dependence. This injects domain knowledge in the model and ensures reliable extrapolations at high temperatures, for instance in conditions pertinent for exoplanetary magma oceans. Other equations can be used~\cite[e.g. see][]{lelosq2021,lelosq2023b}. Here we chose the VFT equation as it is simple and has proven its usefulness in viscosity modeling~\cite{russell2022}. 

Finally, we tested a Gaussian process (GP) model. A GP is a collection of random variables, any finite number of which have consistent joint Gaussian distributions~\cite{rasmussen2006}. In other words, a GP places a probability density over functions. Here, we are interested in placing a probability density over the function $G$ in equation~\ref{eq:funcform}. We do not know this function, but data can allow its probability density to be constrained. A GP is fully defined by its mean $m(x)$ and covariance (a.k.a. kernel) $k(x,x')$ functions:

\begin{equation}
    f(x) \sim GP(m(x),k(x,x')) \,.
    \label{eq:gp}
\end{equation}

$k(x,x')$ choice is critical as it embeds the mathematical formalism to describe how smooth and how variable is the expected function. For instance, a linear kernel implies that the target function is linear. For placing a probability distribution over non-linear functions, one will choose a non-linear kernel such as the squared exponential. Kernels can be multiplied and added, to create complex kernel functions that can reproduce very different shapes of functions. $k(x,x')$ usually has hyper-parameters that will control the behavior of the function (smoothness, periodicity, etc.) and that can be adjusted by gradient-based optimisation. For details regarding kernel design and selection, see~\cite[][]{rasmussen2006}.

The mean function usually is chosen as a constant value equal to 0 in most applications. This is because the mean function is directly updated as we perform predictions on new points $x^*$. It thus usually does not have a strong influence on posterior predictions at those new values $x^*$. However, in the present case, we are dealing with a rather sparse data coverage of a large multidimentional problem. We further may want to extrapolate predictions at the high temperatures typical of exoplanetary magma oceans (2300-3000 K). In such a case, setting $m(x) = 0$ is not the best choice because upon extrapolating, the GP will tend toward the initial value of $m(x)$, thus toward 0. To avoid this, prior information about the problem can be provided in the mean function, in the form of known analytical equations or of a model~\cite{rasmussen2006,ziatdinov2021,zhang2022a}. Here, we use the previously described Greybox ANN as the GP mean function. The Greybox ANN predictions are prior estimates of $\log_{10} \eta$ given $T$, $P$ and $X$ that are updated by the GP at inference time using the kernel function $k(x,x')$ and the data, in order to provide posterior values of $\log_{10} \eta$ at the desired $P$-$T$-$X$ conditions.

The interest of GPs in this work is twofold: first, using appropriate kernels and mean functions, we ensure model smoothness and can leverage domain knowledge, and, second, the GP formalism allows calculating uncertainties on posterior predictions. Uncertainties can be obtain from artificial neural networks using methods such as Monte Carlo Dropout and Conformal predictions~\cite[e.g. see][]{lelosq2023b}, but those are not necessarily straightforward to implement. GP predictions are rooted in Bayesian statistics and the mathematical formalism is well established. GP thus can provide reliable prediction uncertainties, an important point for evaluating the quality of predictions.

\subsection{Hyperparameter tuning and model training}

To tune the hyperparameters of the models, we monitored the root-mean-squared errors (RMSEs) on the training and validation data subsets. The hyperparameters of the LR, SVM and RF models were tuned using Scikit-Optimize, with a Bayesian optimization algorithm. For the ANN model, manual tests and a grid search indicated that a simple yet effective architecture is a network composed of two layers of 200 ReLU activation units~\cite{glorot2011}. The Greybox ANN was implemented using PyTorch~\cite{paszke2019}, with the same architecture as the ANN model but GELU activation functions~\cite{hendrycks2020}, as in i-Melt~\cite{lelosq2023b}. During training, early stopping and dropout~\cite{srivastava2014} were adopted to prevent overfitting. The GP model is implemented using GPyTorch~\cite{gardner2021}. After testing different kernels, we selected a Matérn 5/2 kernel with 14 different lengthscales. The GP mean function is a pre-trained Greybox ANN model. 

Models with optimal hyperparameters were trained on the combined training and validation data subsets, except for the Greybox ANN because early stopping uses the validation dataset during training. For the Greybox ANN and the GP, training may not always proceed well due to the stochasticity in artificial neural network training. This problem is solved by performing 100 different training runs and then selecting only the models for which the metrics were the lowest for the Greybox ANN. The library \textit{gpvisc}~\cite{lelosq2024} allows using trained Greybox ANN and GP models to perform predictions in new situations via Python coding or a more convenient GUI interface. 

\section{Results}

\subsection{Performance evaluation}

Over the very broad range of compositions we investigate (\textbf{Fig.~\ref{fig:dataset}}), the different models predict the viscosity of melts with variable levels of performance. \textbf{Table~\ref{tab:tab1}} reports the RMSE, mean absolute error (MAE), and coefficient of regression ($R^2$) for the different methods and data subsets. Errors on the test data subset are indicative of predictive errors because the models did not see melt compositions included in this data subset. We further report the metrics for the test data subset excluding data points from SciGlass (Test$^*$ column) because those are know to be of variable quality \citep[e.g.][]{wu2015}. Our hand-curated database is of better quality, and error metrics on it may be more indicative of true model errors.

\begin{table}[h]
\centering
\begin{tabularx}{\textwidth}{ X|c|c|c|c} 
\toprule
Model & Metric & Training-Validation & Test & Test$^{*}$\\
\midrule
Linear regression (LR) & RMSE & 1.58 & 1.58 & 1.51\\
& MAE & 1.08 & 1.09 & 0.91\\
& R\textsuperscript{2} & 0.873 & 0.872 & 0.903\\
\midrule
Support Vector Machine (SVM) & RMSE & 0.46 & 0.53 & 0.36\\
& MAE & 0.09 & 0.15 & 0.13\\
& R\textsuperscript{2} & 0.989 & 0.986 & 0.994\\
\midrule
Random Forest (RF) & RMSE & 0.19 & 0.54 & 0.37\\
& MAE & 0.05 & 0.14 & 0.13\\
& R\textsuperscript{2} & 0.998 & 0.985 & 0.994\\
\midrule
Artificial Neural Network (ANN) & RMSE & 0.33 & 0.43 & 0.33\\
& MAE & 0.14 & 0.17 & 0.15\\
& R\textsuperscript{2} & 0.994 & 0.990 & 0.995\\
\midrule
Greybox Artificial Neural Network (Greybox ANN) & RMSE & 0.42 - 0.55 & 0.48 & 0.33\\
& MAE & 0.18 - 0.21 & 0.19 & 0.18\\
& R\textsuperscript{2} & 0.991 - 0.984 & 0.988 & 0.995\\
\midrule
Gaussian process (GP) & RMSE & 0.35 & 0.44 & 0.32\\
  & MAE & 0.12 & 0.15 & 0.14\\
& R\textsuperscript{2} & 0.994 & 0.990 & 0.996\\
\bottomrule
\end{tabularx}
\caption{Root Mean Squared Error (RMSE), Median Absolute Error (MAE) and coefficient of determination $R^2$ of the different models. For final training of the algorithms after hyperparameter optimization, the training and validation data subsets were joined. The RMSE on the joined training-validation dataset is shown here, except for the Greybox ANN model. For this model, we always monitor the metrics on the validation dataset to perform early stopping and avoid overfitting, such that we report the RMSE on the three data subsets. The second Test$^{*}$ column indicates the metrics for the hand-curated part of the test data subset, excluding SciGlass values.}
\label{tab:tab1}
\end{table}

Metrics indicate that the LR model performs poorly, with RMSE higher than 1 $\log_{10} Pa\cdot s$ (\textbf{Table~\ref{tab:tab1}}). This is expected because melt viscosity depends non-linearly on $T$, $P$ and $X$. As a result, non linear models such as SVM and RF perform better, with test RMSE of 0.53 and 0.54 $\log_{10} Pa\cdot s$, respectively. RF has an issue though: the training RMSE is significantly lower than the test RMSE. This indicates that RF overfits its training data, and thus its predictions may not be reliable.

The ANN performs very well with minimal overfitting as indicated by close training-validation and testing RMSEs, respectively of 0.33 and 0.43 $\log_{10} Pa\cdot s$ (test error of 0.33 excluding SciGlass data). This indicates that artificial neural networks are a method of choice for the problem at hand; a result that is not unexpected when considering their success in previous studies~\cite{lelosq2021,langhammer2022,lelosq2023b,cassar2023}.

The Greybox ANN model performs well with minimal overfitting, as indicated by good metrics (\textbf{Table~\ref{tab:tab1}}) slightly higher than those of the blackbox ANN. This is probably due to a slight decrease in model flexibility resulting from the enforcement of the $T$ versus $\log_{10} \eta$ relationship through equation~\ref{eq:TVF}. However, this apparent disadvantage actually is a strength: the Greybox ANN model will always provide physically-realistic viscosity predictions as a function of temperature, even in the high temperature range of magma oceans.

The GP model, based on the combination of the Greybox ANN with a Gaussian process, shows very good predictive performance (\textbf{Table~\ref{tab:tab1}}), with a test dataset RMSE of 0.44 $\log_{10} Pa\cdot s$ (0.32 $\log_{10} Pa\cdot s$ when excluding SciGlass data). GP overfits very slightly its training dataset (RMSE = 0.34) but this remains acceptable. The metrics indicate that, overall, this model outperforms the others.

\subsection{Model selection}

\begin{figure}[ht]
    \centering
    \includegraphics[width=\textwidth]{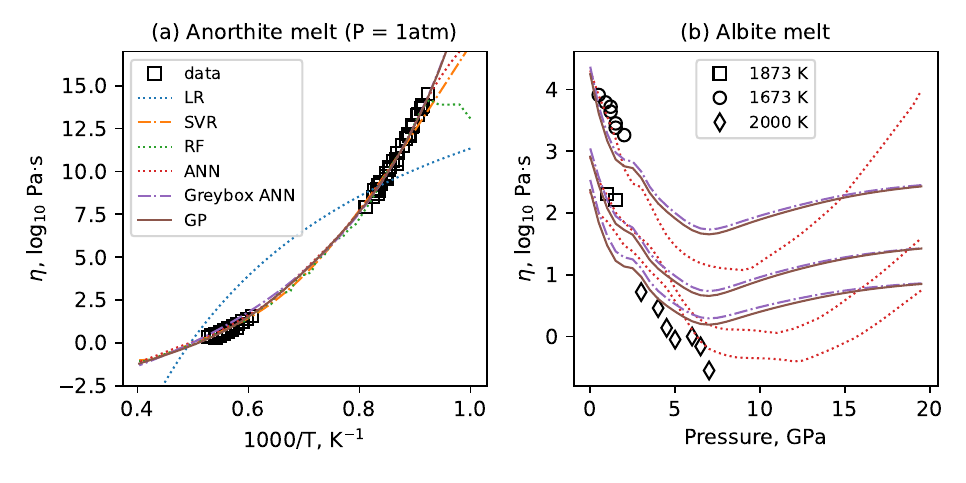}
    \caption{Viscosity against reciprocal temperature of an anorthite melt (a), and against pressure for albite melt (b). Symbols are measurements from the hand-held database \cite[][see references reported herein]{ferraina2024} 
    and lines are model predictions (see legend). Only the predictions of the three best models are represented in (b).}
    \label{fig:tp}
\end{figure}

A closer inspection of the predictions performed by the models reveals that some models are more appropriate for the problem at hand. In particular, predicted viscosities should be continuous functions of intensive variables; that is, pressure and temperature. To evaluate this, melt viscosity is first represented as a function of $T$ for a given melt composition (Fig.~\ref{fig:tp}). In this representation, data tend to cover specific regions (Fig.~\ref{fig:dataset}) as most melts crystallize at $\eta$ comprised between $\approx10^5$ and $\approx10^8$ Pa$\cdot$s. This test confirms that the LR model is not appropriate (Fig.~\ref{fig:tp}a). SVM provides smooth predictions of viscosity, but with small deviations at $\eta > 10^5$. RF only provides sensible results in regions where data are present, its predictions being not smooth. The RF model thus is not appropriate for the present problem, at least in its current simple implementation (more complex implementations are out of the scope of this work).

The ANN, Greybox ANN and GP models all present smooth, realistic predictions of $\eta$ as a function of $T$, even outside of the data calibration range (Fig.~\ref{fig:tp}a), and, indeed as a function of $P$ and composition. The RMSE between calculations and measurements at high pressure is of 0.58 for the GP and of 0.62 for the Greybox ANN (all high pressure data considered). In figure~\ref{fig:tp}b, we show examples of the performance of these models for the prediction of the viscosity of albite melt with pressure at 1673, 1873 and 2000 K. In the range where data are present, the three models yield consistent, smooth predictions. GP predictions are a slight refinement of the Greybox ANN predictions. Above 7 GPa where data are not present for albite melt, the ANN and GP/Greybox ANN models predict diverging values. This reflects the lack of data: models are extrapolating outside their training range and their predictions should be taken with caution (see Discussion).

Among tested models, the Greybox ANN and the GP are favored due to their predictive power (Table~\ref{tab:tab1}). Besides, both embed domain knowledge through the use of equation~\ref{eq:TVF}, this improving the robustness of their predictions. In general, the GP model should be favored as its metrics are slightly better and as it provides uncertainties on predictions. However, in some rare cases, we observed that the Greybox ANN model offered more physically-sound predictions, for instance when asking to predict $\eta$ at high temperature for water-rich melts (see example notebooks at \cite{lelosq2024}). Another case of preference for using the Greybox ANN model is when there is a critical need for speed. Indeed, predictions using this model are faster and more memory efficient than those performed using the GP model. Performing inference using the GP on 10 data points requires $\sim$294 milliseconds on a 13th Gen Intel\textregistered~Core\texttrademark~i9-13900K CPU and $\sim$12 milliseconds on a NVIDIA RTX A4500 GPU. Using the Greybox ANN, the same query is performed in $\sim$524 microseconds on CPU and $\sim$282 microseconds on GPU. GP predictions are fast, but the Greybox ANN model is clearly faster. Therefore, in very intensive fluid dynamic calculations that do not require predictions with uncertainties, it could be desirable to use the Greybox ANN model. The \textit{gpvisc} library \cite{lelosq2024} allows easily getting predictions from both models. For the following analysis in this paper, we will use the GP model.

\subsection{High pressure predictions}

\begin{figure}[ht]
    \centering
    \includegraphics[width=\textwidth]{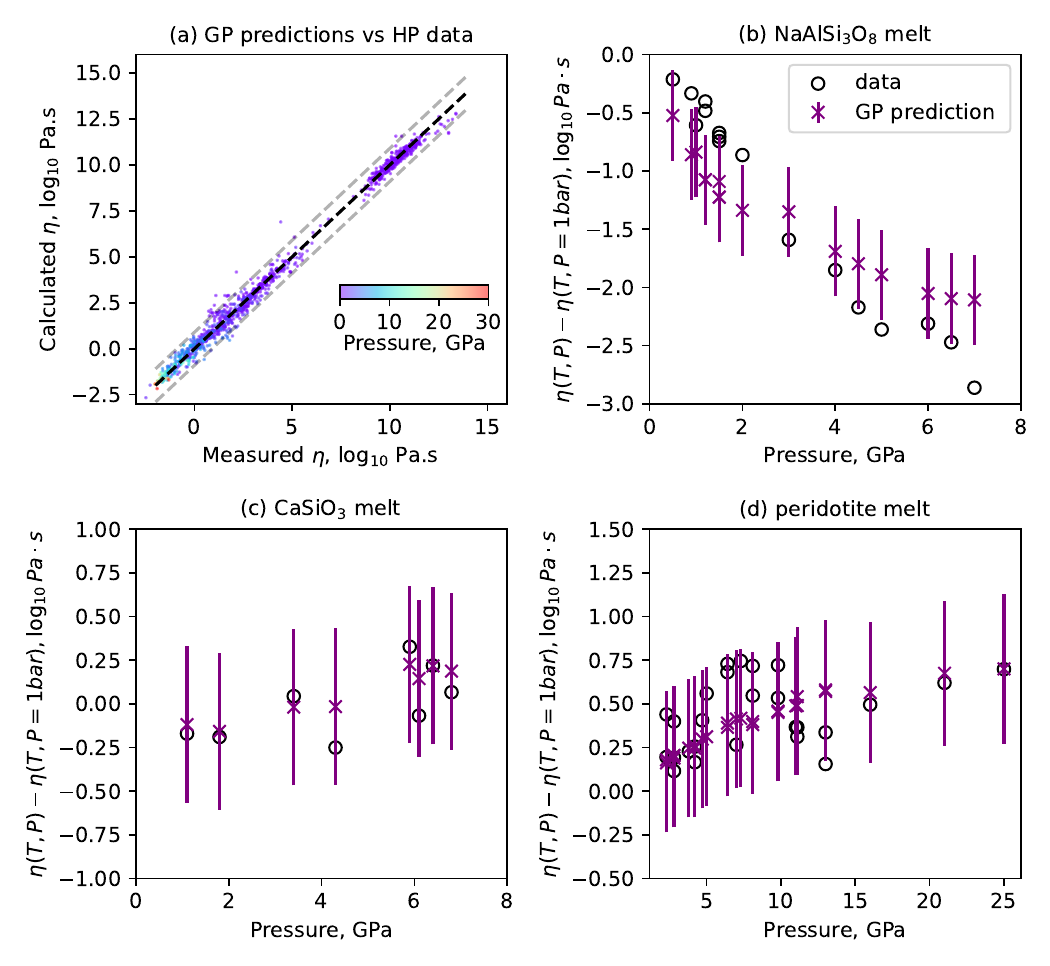}
    \caption{(a) Calculated viscosity with the GP model versus measurements. The black dashed line represent the 1:1 correspondence, and the grey dashed lines the average 95 \% confidence interval ($\pm 0.9$). (b), (c) and (d) Comparison between Gaussian process (GP) viscosity predictions and data available in \cite{ferraina2024} for albite NaAlSi$_3$O$_8$, wollastonite CaSiO$_3$ and peridotite melts. The viscosity at high pressure and high temperature ($\eta (T,P)$) is represented relative to that at the same temperature and 1 bar ($\eta (T,P=1\, bar)$, calculated from an interpolative fit of experimental data with the Vogel-Tammann-Fulcher equation). 
    }
    \label{fig:pressure}
\end{figure}

We now explore in more detail the high pressure predictions using the GP model. Overall, the model reproduces well (within half a log unit) existing high and low melt viscosities data (Fig.~\ref{fig:pressure}a), including those of polymerized melts such as albite up to 7 GPa (Fig.~\ref{fig:pressure}b) as well as those of ultramafic compositions of significance for magma oceans such as CaSiO$_3$ and peridotite up to 30 GPa (Fig.~\ref{fig:pressure}b,c).

However, despite such promising results, we should bear in mind that the high pressure dataset is small and sparse in terms of pressure and compositional coverage. The good predictive performance of the GP model for a composition such as peridotite does not imply that the model can predict viscosity across a range of melt compositions up to 30 GPa.

\subsection{Limitations}

\begin{figure}[ht]
    \centering
    \includegraphics[width=\textwidth]{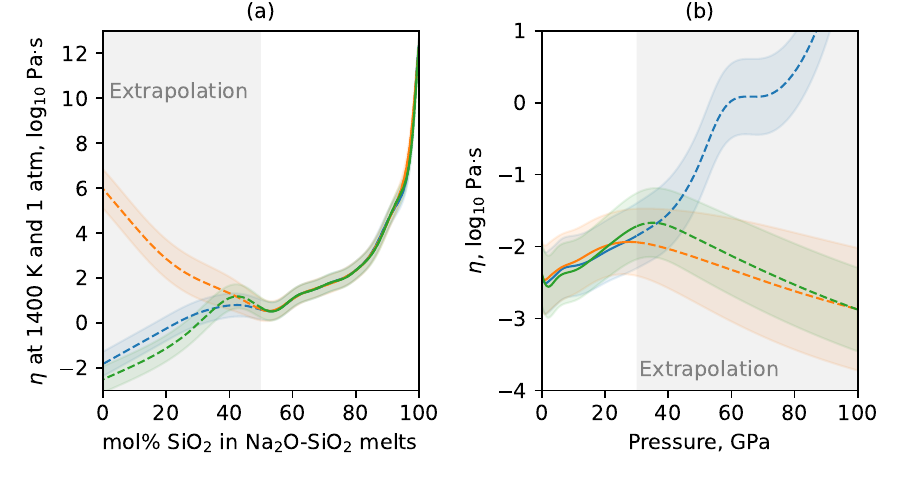}
    \caption{(a) Viscosity in Na$_2$O-SiO$_2$ melts as a function of the silica content. (b) Viscosity at 3000 K as a function of temperature for peridotite melt. Lines are predictions from three GP models; solid lines indicate that the models worked in interpolative regime, dashed lines indicate extrapolations. The grey shaded region indicates the region in which no data was available.}
    \label{fig:extrapolation}
\end{figure}

 One main limitation of ML models is that they are interpolative by nature. Therefore, when asked to predict values for inputs far outside the range of the training data, predictions may be erroneous. At 1 bar, it is unlikely that users ask the Greybox ANN and GP models to extrapolate because the dataset covers a very broad compositional range and eq.~\ref{eq:TVF} constrains the temperature dependence of melt viscosity. However, compositions outside the range of training data may still be queried. To illustrate what happens in such a case, we report in Fig.~\ref{fig:extrapolation}a the viscosity of sodium silicate melts as a function of their silica concentration. Predictions from three different GP models are reported because different training runs yield slightly different models (in particular, slightly different Greybox ANNs), and we can leverage this stochasticity to test the robustness of model predictions. The three GP models have comparable test RMSE values, equal to or lower than 0.45. In figure~\ref{fig:extrapolation}a, above 50 mol\% silica, where data are available, all models yield similar predictions within uncertainties. Below 50 mol\%, their predictions largely diverge. This indicates that the models are providing predictions for compositions outside their training range; they are working in an extrapolative regim and their predictions become unreliable.

We perform the same test at high pressure for a peridotite melt at T = 3000 K (Fig.~\ref{fig:extrapolation}b). Below 30 GPa, predictions from all models are comparable but above 30 GPa, they are distinct. Again, this indicates that the GP models are providing predictions for pressures outside the training range for this composition.
 
Those tests indicate that predictions from models such as the Greybox ANN and GP become unreliable in the extrapolative regim (i.e., for queries outside the range covered by the training dataset). It is not a problem with temperature because the $\log_{10} \eta$ versus $T$ behavior is constrained by equation~\ref{eq:TVF}, but it can be for $X$ and $P$ (Fig.~\ref{fig:extrapolation}). In particular, at high pressure, given the dataset sparsity, such regime could be easily reached for compositions deviating from those in the training dataset. This could be improved by adding a term accounting for the effect of $P$ in equation~\ref{eq:TVF}. A linear term could be added \cite[e.g.][]{russell2024}, but this will not work in all cases because melt viscosity dependence on pressure changes as a function of composition \cite{bottinga1995a}. This area requires further work to be improved.
 
 In case of doubt, the robustness of the predictions can be tested by querying predictions from a few different GP models (Fig.~\ref{fig:extrapolation}). The \textit{gpvisc} library \cite{lelosq2024} provides three trained Greybox ANN and associated GP models to do so, with examples. The test should be performed in case of doubt, because unfortunately the error bars calculated by the GP model are not informative in the extrapolation domain (Fig.~\ref{fig:extrapolation}) as they do not become very large. This may be inherent to the important constraint the Greybox ANN mean function is placing upon the results of the GP model. We also identify this as an area of improvement for future implementations.

Finally, data from molecular dynamics~\cite[e.g.][]{huang2024} could be used to constrain the ML models in the very high pressure range. We did not perform this step here because we focused on presenting models trained on a database that only contains experimental measurements. However, it seems to be a natural improvement for future versions.

\section{Discussion}

\subsection{Machine learning modeling of liquid viscosity}

Using a new database containing viscosity data on very various melt compositions (from unary SiO$_2$ and H$_2$O to geologic and industrial melts) and at high pressure (up to 30 GPa for peridotite), we find that ML models that perform well are the Greybox ANN and the GP model. On unseen data, their average RMSEs are respectively of 0.48 and 0.44 $\log_{10}$ Pa$\cdot$s (Table~\ref{tab:tab1}). Those values compare well with metrics from published models. Taking for instance as a reference point the Giordano et al.~\cite{giordano2008} model of the viscosity of geological melts, its RMSE on unseen data is of 0.7 $\log_{10}$ Pa$\cdot$s. The more recent artificial neural network model of~\cite{langhammer2022} announces a RMSE on unseen data of 0.45 $\log_{10}$ Pa$\cdot$s. Those two models are limited to geological compositions and do not include the effect of pressure. They thus cannot be directly compared to the estimates proffered by models in the present work. Another reference point may be the generalist (i.e.\ very broad compositional domain) GlassNet machine learning model~\cite{cassar2023}. It uses a multitask blackbox artificial neural network to predict various properties, including glass-forming melt viscosity. Applied to the present dataset at room pressure, the RMSE between measurements and GlassNet predictions is equal to 0.95 $\log_{10}$ Pa$\cdot$s. Overall, the Greybox ANN and GP models both either match or outperform existing models while allowing predictions for an extremely wide range of temperature, pressure and compositional scenarios.

The GP model is slightly more accurate than the Greybox ANN model and has the advantage of providing uncertainties on predictions. Inference times are fast, thanks to GPyTorch~\cite{gardner2021}. Such a model can thus be easily incorporated in fluid dynamics simulations without representing a significant computing bottleneck. If faster predictions are required, the Greybox ANN model may be used as its predictions are an order of magnitude faster. This improved speed is at the expanse of slightly less good predictive accuracy in general.

\subsection{Material state and viscosity at surface of dry lava planets}

Using the GP model of melt viscosity, we now will explore the possible properties of the magma ocean on the dayside of a dry hot super Earth, K2-141 b~\cite{malavolta2018,barragan2018}. This planet is of particular interest as it is part of the USP planets studied during JWST Cycle 1 General Observers program~\cite[e.g.][]{dang2021}.

We first calculate K2-141 b surface temperature as a function of longitude. The $\sim$ 48° finite angular size of its star results in a high planetary illumination~\cite{carter2024a} that must be taken into account for the calculation. To do so, we use the analytical and numerical solution described in~\cite{carter2024a} and implemented in~\cite{carter2024}, with the following assumptions: a zero bond albedo~\cite[][]{essack2020}, an average nightside surface temperature of 950 K~\cite[][]{zieba2022}, and the absence of a thick atmosphere as supported by no hotspot offset and no atmospheric heat redistribution~\cite{zieba2022}. A thin rock vapor atmosphere may be present but does not participate in redistributing heat on the planet's surface~\cite{malavolta2018,nguyen2020,zieba2022}. We neglect its influence on the surface temperature calculation.

\begin{figure}[ht]
    \centering
    \includegraphics{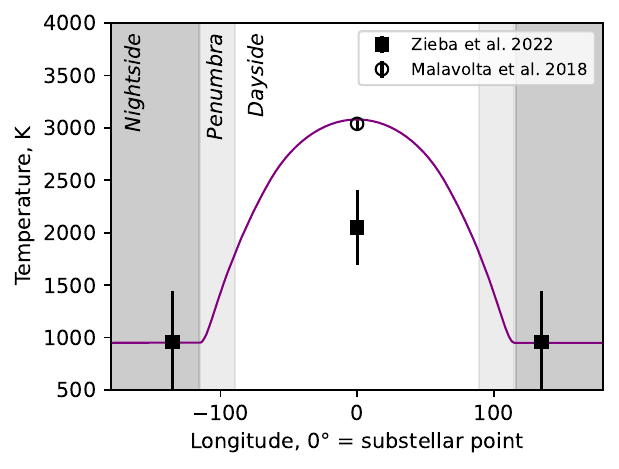}
    \caption{Longitudinal temperature profile at the surface of K2-141 b. See text for details.}
    \label{fig:cancri}
\end{figure}

The longitudinal temperature profile shows a maximum temperature of $\sim 3080$ K at the substellar point (Fig.~\ref{fig:cancri}), in close agreement with~\cite{malavolta2018}. The dayside generally is characterized by temperatures above 1700 K. A penumbral region extends between (-)90° and $\sim$ (-)115°. In this region, temperature drops from $\sim$ 1700 K down to $\sim$ 950 K, the value assumed for the nightside. Given the liquidus temperature of most alumino-silicate rocks, we expect extended melting on the dayside, and magma ocean solidification in the penumbral region.

To examine the effect of composition on magma viscosity, we consider three different compositional scenarios for the magma ocean: Bulk Silicate Earth (BSE)~\cite{palme2014}, BSE + 20 \% FeO (Fe-rich BSE) and a Type-B CAI composition~\cite{stolper1986}. The first scenario is an Earth-like case, the second assumes a higher Fe/Si for K2-141 b, and the last one is considered given the high dayside temperatures that may promote a shift of the mantle composition toward a refractory endmember due to evaporation \cite[][]{leger2011}. The compositions are approximated in simplified systems containing only SiO$_2$, Al$_2$O$_3$, MgO, CaO, and FeO; we did not include the alkalis or volatiles (H and C). Using FactSage, we calculated the fractions of crystals, gas and melts as a function of temperature for the three different scenarios, at atmospheric pressures P\textsubscript{atm} of 10\textsuperscript{-4}, 10\textsuperscript{-1} and 1 bar.

We first explore the possible longitudinal profile of evaporation/condensation. FactSage calculations provide equilibrium fractions of gas, melt and solids at high temperatures. Using those, we neglect phenomenons that could bring melt out of equilibrium with the gas, such as magma ocean convection and atmosphere evaporation and circulation. Despite this, this approach can provide pieces of understanding regarding the possibility of evaporation and condensation of rock vapor from the magma ocean, assuming local equilibrium (in terms of space and time). Given the longitudinal temperature profile, we calculated the longitude at which the system is comprised of greater than 1\% of gas (Table~\ref{tab:tab2}). At P\textsubscript{atm} = 1 bar, temperatures are not high enough to generate significant outgassing for the present alkali-free compositions. At 10\textsuperscript{-1} bar, outgassing starts at longitudes lower than $\sim$ 37° for the BSE and Fe-rich BSE case, and is completely inhibited for the CAI case. At 10\textsuperscript{-4} bar, outgassing should occurs at longitudes lower than 70°-80° in the three compositional scenarios.  

\begin{table}
    \centering
    \begin{tabular}{ cccc }  
    \toprule
    P\textsubscript{atm} (bar) & BSE & Fe-rich BSE & CAI \\ 
    \midrule
    10\textsuperscript{-4}  & 77 & 78 & 72\\ 
    10\textsuperscript{-1}  & 36 & 37 & 0 \\ 
    1                       & 0  & 0  & 0 \\ 
    \bottomrule
    \end{tabular}
    \caption{Longitudes (degrees) at which the temperature allows for 1\% gas and 99\% melt at equilibrium for three compositional scenarios and three different atmospheric pressures P\textsubscript{atm}.} 
    \label{tab:tab2}
\end{table}

\begin{figure}[h]
    \centering
    \includegraphics{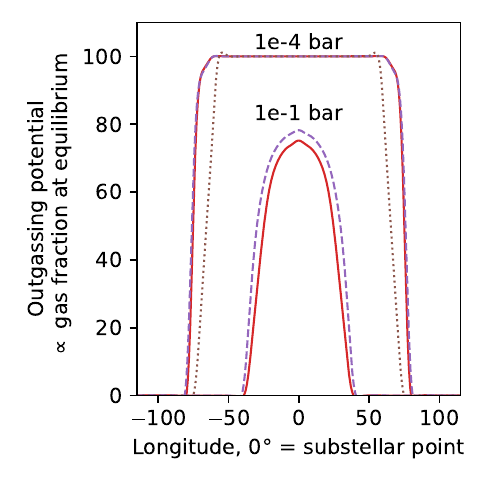}
    \caption{Magma ocean outgassing potential as a function of longitude at the surface of K2-141 b for the BSE (solid lines), Fe-rich BSE (dashed lines) and CAI (dotted lines) compositional scenarios at two different pressures.}
    \label{fig:outgas}
\end{figure}

Another way of visualizing those results is to use the gas fraction obtained for a given mixture of gas and melt at equilibrium at different P\textsubscript{atm} and temperatures as a proxy for the outgassing potential. This assumes that high gas fraction in the equilibrium closed-system case will imply more vigourous outgassing at the surface of the dynamic magma ocean. Reporting the equilibrium gas fraction as a function of longitude shows that a 10\textsuperscript{-1} bar atmosphere is expected around the substellar point (Fig.~\ref{fig:outgas}). A lower pressure atmosphere is expected at longitudes above 37°, and up to $\sim$ 78° for a P\textsubscript{atm} = 10\textsuperscript{-4}. 

This simple evaluation disregards the presence of volatile and/or alkali elements in the magma. Oxygen fugacity $f$O$_2$ and temperature also may have important effects on total atmosphere pressure, as shown by the equilibrium models of \cite{seidler2024}. The latter study finds that near 3000~K, surface pressures total 10$^{-2}$ bar at the iron-wüstite (IW) buffer, increasing up to $\sim$3.5 bar under reducing conditions ($\Delta$IW = -6). Another limitation is the absence of consideration for atmosphere circulation, which could imply evaporation near the sub-stellar point and a gradual condensation as gas moves toward the penumbral region~\cite{kite2016}. Considering such complexities is outside the scope of this work. Despite this, the present analysis agree with, and corroborate the results from the more complete study of~\cite{zieba2022}. Their calculations suggest a rock-vapor atmosphere of 10\textsuperscript{-1} bar up to $\sim$ 40° longitude at K2-141 b surface, its pressure decreasing when going toward the penumbral region. Assuming a lateral atmospheric current between the substellar point and the penumbral region, outgassing is thus expected in the 40° region around the substellar point, and rock vapor condensation should occur in the lower pressure regions, between 40° and 90°.

\begin{figure}[ht]
    \centering
    \includegraphics{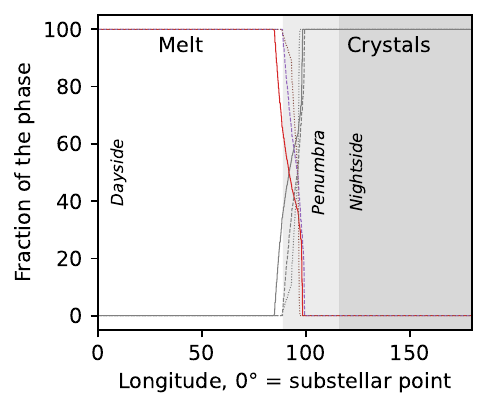}
    \caption{Percent of melt and crystals at surface as a function of longitude for the BSE (solid lines), Fe-rich BSE (dashed lines) and CAI (dotted lines) compositional scenarios. Colored lines are melt fractions and grey lines are crystal fractions.}
    \label{fig:cancri_phases}
\end{figure}

We now consider the fractions of melt and crystals the magma ocean contains at surface (Fig.~\ref{fig:cancri_phases}). In all three cases, the magma ocean is fully molten at the dayside surface (Fig.~\ref{fig:cancri_phases}). Differences only are visible close to the shores, in the penumbral region. The refractory CAI composition presents the sharpest transition from fully molten to solid, and the BSE has the largest transition region from melt to solid. However, for all compositions, we expect a rapid solidification of the magma ocean at the beginning of the penumbral region.

Given variations of melt composition and crystal/liquid fractions, we then calculate magma viscosity using the GP model for the liquid phase. To calculate the effect of crystals, we use the model of~\cite{costa2009} for the liquid-rich magmatic mixture (fraction of crystals $\phi$ lower than 65 \%). It includes parameters that account for crystal size and shape distributions as well as maximum packing fraction. For those parameters, we used values determined for peridotite compositions provided in \cite{costa2009}. For crystal-rich magmatic mixtures, we use the model from~\cite{scott2006} (at $\phi $ > 65 \%), assuming a reference viscosity for the solid mantle of $10^{18} \log_{10}$ Pa$\cdot$s at 1273 K.

\begin{figure}[ht]
    \centering
    \includegraphics{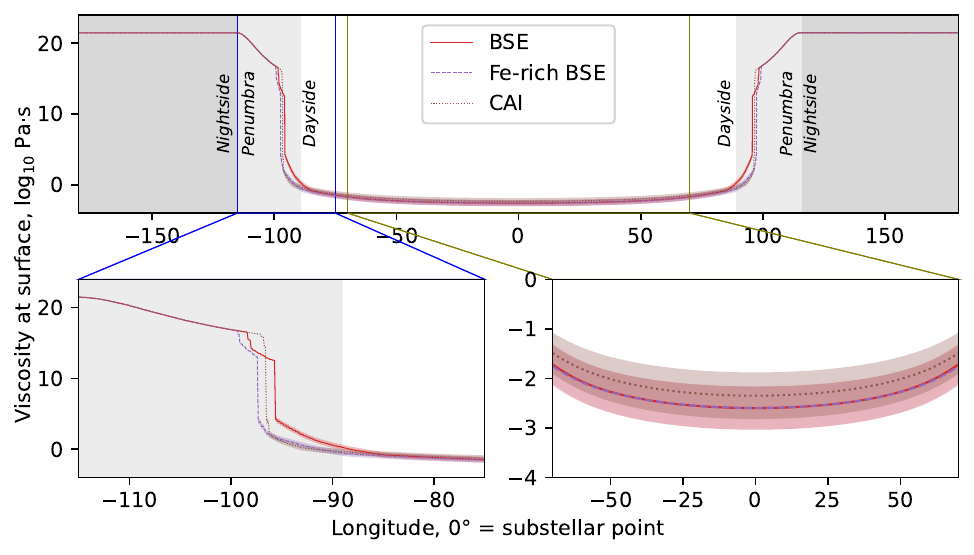}
    \caption{
         Magma viscosity at surface given longitude, calculated assuming a Earth-like composition (Bulk Silicate Earth, BSE), a Fe-rich composition (Fe-BSE) and a refractory composition (Type B CAI, see~\cite{stolper1986}). See text for details. Shaded regions on the side of the lines are 95 \% confidence intervals.}
    \label{fig:cancri_visco}
\end{figure}

We observe little variations in magma ocean viscosity between the different scenarios (Fig.~\ref*{fig:cancri_visco}). In most of the dayside, magma viscosity is lower than $10^{-1}$ Pa$\cdot$s. At the substellar point, magma viscosity reaches values of $\sim 10^{-2.3}$ and $\sim 10^{-2.6}$ Pa$\cdot$s for the CAI and BSE/Fe-BSE cases, respectively. Melt composition thus has a small impact ($\sim 0.3$ log units) on the viscosity of the fully molten magma. When going toward the nightside, the earlier appearance of crystals in the BSE case (Fig.\ref*{fig:cancri_phases}) implies that magma viscosity increases sooner when transitioning from the dayside to the nightside. However, the difference is only of a few degrees compared to the Fe-rich BSE and the CAI scenarios. 

Overall, this analysis highlights two points:
\begin{itemize}
    \item in the high- to very high temperature regimes characteristic of magma oceans on hot super-Earths, temperature exerts the zeroth-order control on magma viscosity, erasing the influence of other factors usually critical in Earth-like volcanic systems (e.g. melt composition);
    \item when temperature decreases, crystallization dictates the rheology of magma, due to combined effects of the presence of solid particles and of the changes it imparts on the residual melt composition. This effect will be important on the shores of magma oceans, but possibly also at depth as crystallization may be favored by the increasing pressure. We do not explore this here because it requires an internal model of the planetary structure that is beyond the scope of this paper.
\end{itemize}

\subsection{K2-141 b: nightside temperature}

On the nightside, we expect a solid surface (Fig.~\ref{fig:cancri_phases}) with a very high viscosity (Fig.~\ref{fig:cancri_visco}). K2-141 b has a very tenuous atmosphere, unlikely to redistribute heat toward the nightside~\cite{zieba2022}. Therefore, an internal geothermal heat flow will drive heat redistribution between the dayside and nightside. The nightside surface temperature $T_{s}$ will be controlled by a balance between the geothermal heat flux $Q_{geo}$ and the radiative surface heat flux $Q_{rad} = \sigma _{B} T_{s}^{4}$, with $\sigma _{B}$ the Stefan-Boltzmann constant. $Q_{geo}$ can be sustained either by horizontal transport of heat from the dayside to the nightside by a thin magma ocean under a lid, or by vertical transport from the interior to the surface by the rocky mantle (potentially partially molten). In the case of horizontal convection, we have~\cite{hughes2008}:
\begin{equation}
    Q_{geo} = k\frac{\Delta T}{L}R_a^{1/5} \:,
    \label{eq:horiconv}
\end{equation}
    with $k$ the average thermal conductivity, $\Delta T = T_{s} - T_{i}$ with $T_{i}$ the average internal temperature of the convecting layer and $L$ its height, and $R_a$ the Rayleigh number equal to
\begin{equation}
    R_a = \frac{\rho g \alpha \Delta T L^3}{D\eta} \:,
    \label{eq:Ra}        
\end{equation}
where $g$ is the gravity acceleration, and $\rho$, $\alpha$, $D$ and $\eta$ the average values of density, thermal expansivity, thermal diffusivity and viscosity of the material composing the convecting layer. For vertical convection, two cases can be distinguished: hard and soft convection. In the case of a mantle containing a large fraction of melt (below the rheological threshold at around 60 vol\% solids), $R_a$ will be very high ($\approx$10$^{30}$ for an Earth-sized planet) and hard convection should be considered, such that we have~\cite{solomatov2000}:
\begin{equation}
    Q_{geo} = 0.22 k\frac{\Delta T}{L}R_a^{2/7}P_r^{-1/7}\lambda^{-3/7} \:,
    \label{eq:hardconv}    
\end{equation}
where $P_r = \frac{D}{\rho\eta}$ is the Prandtl number and $\lambda$ is mean flow aspect ratio. We assume $\lambda = 1$, following~\cite{solomatov2000}. Now, if the mantle is mostly composed of crystals, it will undergo the rheological transition and $R_a$ will decrease significantly. Compared to the fully molten magma ocean, we expect a decrease by approximately 15 orders of magnitude as viscosity drastically increases (Fig.~\ref{fig:cancri_visco}). Soft convection should then be considered and the heat flux can be expressed as~\cite{solomatov2000}:
\begin{equation}
    Q_{geo} = 0.089 k\frac{\Delta T}{L}R_a^{1/3} \:.
    \label{eq:softconv}    
\end{equation}
By equating $Q_{geo}$ with $Q_{rad}$, we can calculate the nightside surface temperature by iteration. We performed this for two cases: horizontal and hard convection. We neglect the case of soft convection as it is valid for crystal-rich cases (above the rheological threshold). Actually, in details, equation~\ref{eq:softconv} does not really apply for solid mantle convection and another functional form should be adopted~\cite{solomatov2015}. Here, we neglect this and use the hard convection case to provide an upper limit on $T_s$ for crystal-rich cases.

\begin{figure}
    \centering
    \begin{subfigure}[b]{0.45\textwidth}
        \includegraphics[width=\textwidth]{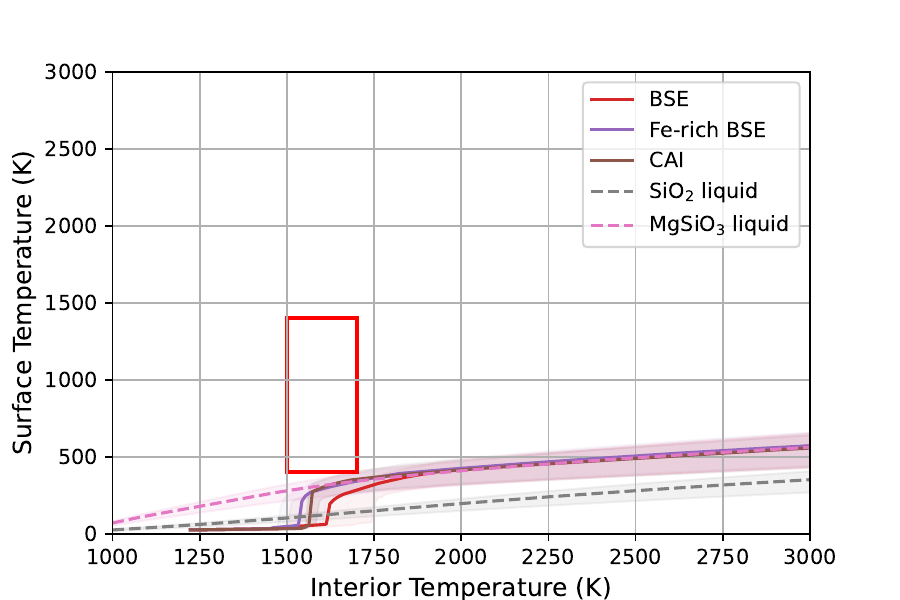}
        \caption{}
        \label{fig:hori_conv}
    \end{subfigure}
    \begin{subfigure}[b]{0.45\textwidth}
        \includegraphics[width=\textwidth]{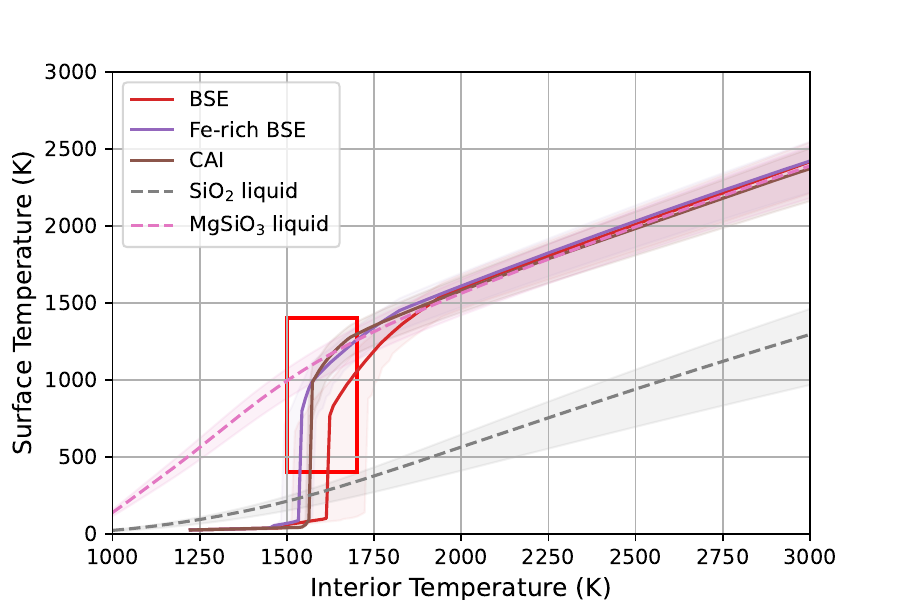}
        \caption{}
        \label{fig:hard_conv}
    \end{subfigure}
    \caption{Average nightside surface temperature as a function of average nightside internal temperature of the convecting layer for two cases: horizontal (a, eq.~\ref{eq:horiconv}) and hard (b, eq.~\ref{eq:hardconv}) convections. The red shaded area delimits the temperature range estimated for K2-141 b~\cite{zieba2022}. The curves are the mean values of 200 models, calculated by sampling the parameter distributions described in the text. Shaded areas around those curves represent the 95\% confidence intervals of model predictions.}
    \label{fig:nightside_T}
\end{figure}

Figure~\ref{fig:nightside_T} shows the result of the calculation for a convecting layer of 1000 km, and for the three cases BSE, Fe-rich BSE and Type B CAI. For the calculations, we used $\rho \sim \mathcal{N}(4000,\,400^{2})$, $\alpha \sim \mathcal{N}(6e-5,\,6e-6^{2})$, $D \sim \mathcal{N}(5e-7,\,2e-7^{2})$, $C_p \sim \mathcal{N}(1500,\,150^{2})$ following experimental values and calculations from~\cite{eriksson2003,gibert2003,suzuki1998,richet1985,lelosq2023b}. We neglect variations of those properties with melt composition as their influence is minimal because they do not vary by more than one order of magnitude. On the contrary, the convecting layer viscosity can vary by up to 27 orders of magnitude as a function of temperature and magma crystal content (Fig.~\ref{fig:cancri_visco}). In particular, the influence of crystallization on melt viscosity is paramount, and unfortunately one uncertainty remains: the volumetric fraction of crystals at which the rheological threshold, $\phi_c$, occurs. $\phi_c$ is usually taken to be at a solid fraction of 0.60~\cite{kolzenburg2022}. Below $\phi_c$, the convecting layer has the rheological behavior of a crystal-bearing liquid. Calculated magma viscosities are lower than $10^6$ Pa$\cdot$s and $R_a$ will be above $10^{25}$. However, above $\phi_c = 0.60$, the rheological behavior changes. The convecting layer adopts the rheological behavior of a partly-molten solid, magma viscosity largely increases and $R_a$ too. $\phi_c$ value is thus critical. It varies significantly with crystal size and shape~\cite{kolzenburg2022}. Here, to take into account of the uncertainty surrounding ~$\phi_c$ value in the different scenarios, we assume for $\phi_c$ a normal distribution with a mean of 0.6 and a standard deviation of 0.06 (10\% error uncertainty).

Figure~\ref{fig:hori_conv} reveals that horizontal convection is not efficient in redistributing the thermal energy. If this mode is dominant, we do not expect temperatures above 600 K at the nightside surface of a dry exoplanet such as K2-141 b, even in the presence of a molten magma ocean beneath a lid. Vertical hard convection is more efficient at providing a sustained geothermal flux (Fig.~\ref{fig:hard_conv}). In this case, if the average internal nightside temperature is above $\sim$1600 K, the nightside surface can reach temperatures above 400 K, the apparent lower limit for K2-141 b~\cite{zieba2022}. The $\sim$1600 K interior temperature threshold is easily understandable: at this temperature, the rheological transition occurs and the liquid-rich magma becomes efficient in redistributing heat. The average nightside temperature then starts to closely mirror the average internal nightside temperature.

We finally note that little differences are discernible between the three different magma compositions. The visible difference concerns the onset of the nightside surface temperature rise for a given internal nightside temperature when the latter is around 1600 K. This arises and is due to differences in how the crystallisation sequences between the BSE, Fe-BSE and CAI magma depend on temperature between the melt compositions. In any case, most uncertainties for the determination of the curves shown in figure~\ref{fig:nightside_T} arise from the uncertainty on $\phi_c$. The maximal influence of melt composition can be observed when comparing the endmember cases of fully depolymerized MgSiO$_3$ and fully polymerized SiO$_2$ melts (Fig.~\ref{fig:nightside_T}). While the curve for liquid MgSiO$_3$ closely follows those of BSE, Fe-rich BSE and CAI (except below $\sim$ 1600 K as MgSiO$_3$ is considered fully molten), a fully polymerized, very viscous SiO$_2$ liquid mantle will not redistribute efficiently heat to the surface. 

The present calculations have inherent limitations, such as the absence of any consideration regarding temperature, pressure and associated property gradients. Besides, while horizontal convection appears unlikely to efficiently transport heat, the reality may be more complex as this mode is not well understood in the case of the laterally varying surface heating that applies for USP lava planets~\cite{meier2023}. Despite such limitations, the present calculations yield an important conclusion. For sustaining a nightside temperature above 400 K without atmospheric redistribution, as suggested for K2-141 b~\cite{zieba2022}, a portion of the nightside mantle should be partly molten. From figure~\ref{fig:nightside_T}, we estimate that this partly molten mantle portion should have a melt fraction at or slightly above 0.4.

\section{Conclusion}

In this paper, we presented a new database and its use to train machine learning models to predict the viscosity of water-bearing phospho-alumino-silicate melts as a function of chemical composition, temperature and pressure. Greybox artificial neural networks produce good results. Their predictions can further be refined using a Gaussian process (GP) to achieve good predictive accuracy (RMSE $\sim 0.4 \log_{10}$ Pa$\cdot$s) and further have access to confidence intervals on predictions. In its current implementation, the GP model allows calculating the properties of magmatic melts in a wide variety of scenarios, including (water-rich) magmatic systems in subduction regions, alkali-rich intraplate volcanism, magma oceans, and extra-terrestrial volcanisms involving compositions far from the Earth trends. Glass manufacturing may also use this model to calculate melt viscosity in glass-forming furnaces, with potential applications in furnace or compositional design for ensuring energy efficient and environmentally effective production of glass. Finally, while pressure effects can be reproduced up to 30 GPa for compositions like peridotite, the sparsity of high pressure viscosity data hampers the ability of ML models to find general laws expressing how melt viscosity varies with pressure for any composition. This invites to further work to better constrain the viscosity of silicate melts at pressures above 1 GPa.

Using the new model as well as phase equilibrium and temperature calculations, we evaluate the properties of a magma ocean at the surface of the dry USP exoplanet K2-141 b, given three different compositional scenarios. Assuming a null bond albedo, calculation of surface temperature indicates that K2-141 b dayside is fully molten in all scenarios. On the dayside, the extreme temperatures generally dictate melt viscosity, and other parameters such as melt composition are secondary. Phase calculations indicate that a tenuous atmosphere with a pressure of $10^{-1}$ bar may be present in a 40° around the substellar point, in agreement with~\cite{zieba2022}. When going to higher longitudes, the atmospheric pressure is expected to drop to very low values. When reaching the penumbral region at a longitude of 90°, magma viscosity increases rapidly as magma ocean solification proceeds. The composition of the magma slightly affects the position of the shores, but this effect is limited (a few degrees). In the nightside region, surface is expected to be solid. However, estimated nighside temperatures above 400 K~\cite{zieba2022} suggest that the mantle below the surface is partly molten and feeds the geothermal flux through vertical convection.

\section*{Acknowledgment}

CLL thanks S. Charnoz (IPGP) for various discussions on the present topic and his help in catching the $R_p$ scaling of the illumination in Carter \cite{carter2024} code.

\section*{Funding}
This study was supported by the LabEx UnivEarthS, ANR-10-LABX-0023 and ANR-18-IDEX-0001.

\section*{Author contributions}
Study design: all authors. Database construction: CF, CLL. Machine learning modelling: CLL, CF. Phase diagram calculations: PS. Temperature and fluid dynamics calculations: CEB, CLL. Manuscript draft: CLL. Manuscrit redaction: all authors.

\section*{Competing interests}
Authors declare no competing interests. 

\section*{Materials \& Correspondence}
The hand-held viscosity database is available at \cite{ferraina2024}. The computer code to reproduce the results of this study is available as a Python library at the web address \url{https://github.com/charlesll/gpvisc} and on Zenodo \cite{lelosq2024}. Correspondence can be addressed to the corresponding author.

\bibliography{ms}  
\appendix
\section*{Appendix A: Data splitting}
For performing the training-validation-testing splits, we first separated the dataset in two room pressure and high pressure datasets, because the compositional coverage and the number of data at room and high pressures are very different and skewed towards the former. For room pressure data, we adopted a stratified group splitting approach that ensures that each subset will contain different compositions in relatively balanced proportions~\citep[see][for details]{lelosq2023b}. For high pressure data, we did not use this method because the high pressure dataset contains too few compositions, with only a few data points at different pressures per composition. Considering that the compositional domain will be already well covered and represented in the different room pressure data subsets, and that the aim of the high pressure data is to provide knowledge about the effect of pressure, we decided to split randomly the high pressure data in the training, validation and testing data subsets regardless of melt compositions. To perform the different random splits, we selected optimal random seeds such that the training, validation and test subsets present similar statistical distributions of their features. To do so, we tried 1,500 different random seeds, and selected the 10 ones that ensured the lowest Wasserstein distances of the feature distributions between the various data subsets. Then, among those seeds, we rejected those that resulted in including end-member compositions (e.g., SiO$_2$ or H$_2$O) in the validation or testing data subsets. We then selected an appropriate seed given this constrain and the Wasserstein distances test. This approach ensures the similarity of the data distributions between the different training, validation and test data subsets. Finally, after data splitting, the  high and low pressure parts of the training-validation-testing data subsets were concatenated. 

\end{document}